\documentstyle[preprint,aps,epsf,psfig]{revtex}
\begin{document}

\title{Self-assembly in Lyotropic Chromonic Liquid Crystals}
\author{Prabal K. Maiti, Yves Lansac, M. A. Glaser and N. A. Clark}
\address{Department of Physics and\\
Ferroelectric Liquid Crystal Material Research Center,\\
University of Colorado, Boulder, CO 80309}
\maketitle
\begin{abstract}
We have developed a class of idealized models of chromonic
molecules which are miscible in water, but which can form aggregates
which in turn organize into lyotropic liquid crystal (LC) phases. 
By carrying out Monte Carlo simulation in a binary mixture of model 
chromonic and water molecules, we have studied the effect of concentration
and molecular shape on the nature of resulting 
mesophases. We have also computed the free energy associated with 
the formation of chromonic columnar aggregates  by umbrella sampling. Our free
energy computation helps us to verify the isodesmic behavior which is 
characteristic of chromonic systems. For isodesmic behavior the addition 
of each chromonic molecule to the columnar aggregates is accompanied by a 
constant free energy increment and the net free energy decreases with increasing
column length.
\end{abstract}

\section{Introduction}
Recently there is a growing interest in a new class of lyotropic liquid crystals
(LCs), so called lyotropic chromonic liquid crystals (LCLCs) due to their 
diverse potential application in many areas such as dyes,
drugs, nucleic acids, antibiotics and anti-cancer agents ~\cite{lydon1},. 
Dichroic thin films formed from LCLCs can be used both as alignment 
layers and polarizers for LC cells \cite{oleg}.
These materials are fundamentally different from conventional amphiphilic systems: their
molecules are disc-like or plank-like rather than rod-like, 
rigid rather than flexible and 
aromatic rather than aliphatic. Due to their disc-like molecular shape 
chromonic molecules stack face to face forming columnar aggregates. 
Though there is no clear understanding of the nature of forces responsible 
for this face-to-face stacking, it is generally believed that the $\pi$-$\pi$
interaction of the aromatic cores is the main mechanism of molecular 
face-to-face stacking ~\cite{lydon1}.  
Hydrophilic ionic groups at the periphery of the molecules make them water soluble.  
The two principal chromonic 
mesophases are chromonic $N$ and $M$ phase. The basic structural unit
of both phases is the untilted stack of the molecules. The $N$ phase is 
the nematic phase in which the molecular stacks has orientational order but
no positional ordering and is formed when the concentration is low.
At higher concentration these molecular 
stacks arrange themselves in a hexagonal pattern forming the $M$ phase.  
In recent years, many new chromonic phases has been reported ~\cite{harrison,tiddy} 
which are formed by aqueous cyanine dyes. Some of the liquid-crystalline
aggregates formed by the cyanine dyes are shown in Figure ~\ref{fig1}.

Unlike the conventional 
amphiphilic system, the chromonic system does not have any critical 
micellar concentration (CMC). 
To avoid the unfavorable contact of the hydrophobic part of the molecules with 
the solvent, the molecules start to stack together forming columns even 
in dilute solution. As the 
column length increases, the fraction of the total molecular surface
exposed to the solvent decreases. There is widespread experimental evidence 
that there is no optimum column length or aggregation number like the 
conventional amphiphilic system forming micellar aggregates. The micelle
has a optimum size depending on the packing parameters which in turn depend 
on the ratio of the size of head and tail group \cite{tanford}. 
The chromonic aggregation behavior
is termed as `isodesmic' in contrast to the `non-isodesmic' behavior observed
for simple amphiphilic system. For isodesmic behavior, addition of each 
chromonic molecule to the columnar aggregate is accompanied by a constant
free energy increment. Therefore, with the increase in column length the 
free energy gradually decreases, in contrast with the non-isodesmic
behavior of simple surfactant system, where micelle formation is accompanied
by sharp free energy minima.  
The self-assembly of chromonic molecules is governed by many factors such as 
molecular structure, concentration, temperature, solvent polarity and ionic 
strength. So far there is no clear understanding of 
the relationship between molecular structure and the structure of the supramolecular
aggregates formed by the LCLC materials. Consequently many different aggregate
structures have been proposed based on different experimental results.

In this paper, we have developed idealized molecular models of LCLCs which
capture the overall shape of the LCLC molecules and the nature of their 
interactions in an approximate way. To understand the self-assembly of 
chromonic molecules, we have carried out simulations of binary mixture of
LCLC molecules and water, and have studied the effect of concentration, 
temperature and molecular shape on the nature of the resulting mesophases. 
We have also computed the free energy associated with the addition of each 
chromonic molecule to a columnar aggregate. The method gives us the 
possibility to test the widely accepted isodesmic behavior of the 
chromonic aggregation.

The paper is organized as follows: in section 2 we briefly describe 
our model and the simulation technique. 
The results on the columnar aggregates, in bulk water, formed
by the chromonic molecules are reported in section 3.
We describe
the method of computation of free-energy by umbrella sampling in section 4.
Finally, a summary of the main results and the conclusions drawn from these
are given in section 5.

\section{The Model}
We have modeled chromonic molecules as diamond shaped constituted of $9$ (model 1)
tangent spheres of diameter $\sigma$ bonded together (figure 2).
We also considered another variant of the model where the chromonic molecules
are disk shaped constituted of $7$ (model 2) tangent sphere of diameter $\sigma$ 
bonded together (figure 2). Model 2 is similar to the model considered by 
Edwards et. al. \cite{edwards} with different kind of interactions.
In case of model 1 the $7$ spheres forming the disc are hydrophobic and the 
$2$ spheres at the two ends are hydrophilic. In case of model 2, the $6$ outer
spheres are hydrophilic and the inner sphere at the center of each molecule is
treated as hydrophobic. Water molecule is modeled as a sphere.

Attractive interaction between like particles (water-water, hydrophilic-water,
and hydrophilic-hydrophilic) is modeled 
via a Lennard-Jones (LJ) potential: 

\begin{equation}
V(r) = 4\epsilon\{(\frac{\sigma}{r})^{12} - (\frac{\sigma}{r})^6 \}
\end{equation}

where $r$ is the separating distance between the two interacting particles. 
The parameter $\epsilon$ governs the strength of interaction and $\sigma$
defines a length scale. We have assumed that $\sigma =1$ and $\epsilon=1$
for all interactions. We have taken a cut-off $r_c = 2.5 \sigma$, large enough to 
include excluded volume effects and attractive forces.

Repulsive hydrophobic-hydrophilic and water-hydrophobic interactions are 
modeled by a truncated and shifted LJ potential (WCA potential):

\begin{eqnarray}
V(r) & = & 
4\epsilon\{(\frac{\sigma}{r})^{12} - (\frac{\sigma}{r})^6 \} +\epsilon 
~~~~~~~~~r \le r_c ,\nonumber \\
       & = & 0 ~~~~~~~~~~\it{otherwise}.
\end{eqnarray}

where $r_c = 2^{1/6} \sigma$.

For MC studies using LJ potential it is convenient to introduce reduced units.
In terms  of $m$, $\sigma$ and $\epsilon$, some of the quantities of interest
are density $\rho^* = \rho \sigma^3$, temperature $T^* = k_BT/\epsilon$, and pressure
$P^* = P\sigma^3/\epsilon$.
The simulation has been carried out both in $NVT$ (fixed number of particles
at constant volume and temperature) and $NPT$ fixed number of particles
at constant pressure and temperature) ensemble. 
During each MC step either chromonic or water molecules were chosen
randomly and displaced using Metropolis criteria. The reorientation move was
performed using quaternions ~\cite{allen}.
We have carried out simulations for different concentrations of the
chromonic molecules at different temperatures. The concentration of chromonic is defined as
\begin{equation}
\phi = \frac{xN_c}{xN_c + N_w}
\end{equation}
where $x$ is the number of atoms constituting a single chromonic molecules, $N_c$
and $N_w$ are the number of chromonic and water molecules respectively, $x$ is $9$ for 
model 1 and $7$ for model 2. 

\section{Results}
In order to investigate the spontaneous formation of chromonic aggregates
and their morphology, various choices of initial conditions were used.
Model chromonic and water molecules are initially dispersed
randomly in a $L_x \times L_y \times L_z$ system with
periodic boundary condition. 
Unless specified all of the results shown below are 
for model 1. Instantaneous snapshots of the columnar aggregates formed 
by chromonic molecules are shown in figure \ref{snap1} and figure \ref{snap2}. 
At low concentration the chromonic molecules form short columns and with 
the increase in concentration the length and the number of the columnar aggregates 
increases. At higher concentration they form chain like aggregates.

We have also calculated
various positional and orientational pair correlation functions for the 
chromonic as well as water molecules to characterize the structure of the 
columnar aggregates. 
%$g_{cc}(r)$, $g_{ss}(r)$, 
%$g^r_{cc}(r^{\parallel}, r^{\perp})$, $g^1_{cc}(r^{\parallel}, r^{\perp})$
$g^{\rm cc}(\bf r)$ is the radial 
distribution function of the center of mass of the chromonic molecules and 
is defined as
\begin{equation}
g^{\rm cc}({\bf r}) = {1 \over {\rho N_c}} \left\langle
\sum_{i \neq j} \delta({\bf r} - {\bf r}_{ij}^{\rm cc})
\right\rangle
\end{equation} 
$g^{\rm ss}(\bf r)$ is the same for  solvent molecules (water) and is defined in an 
analogous way. 
$g_1^{\rm cc}({\bf r})$ and $g_2^{\rm cc}({\bf r})$ are the orientational correlation 
functions and are defined as

\begin{equation}
g_1^{\rm cc}({\bf r}) = {1 \over {\rho N g^{\rm cc}({\bf r})}}
\left\langle
\sum_{i \neq j} P_2({\bf u}_i^{\rm cc} \cdot {\bf u}_j^{\rm cc})
\ \delta({\bf r} - {\bf r}_{ij}^{\rm cc}) \right\rangle,
\end{equation}
\begin{equation}
g_1^{\rm cc}({\bf r}) = {1 \over {\rho N g^{\rm cc}({\bf r})}}
\left\langle
\sum_{i \neq j} P_2({\bf pu}_i^{\rm cc} \cdot {\bf pu}_j^{\rm cc})
\ \delta({\bf r} - {\bf r}_{ij}^{\rm cc}) \right\rangle,
\end{equation}

where ${\bf r}_{ij}^{\rm cc}$ is the relative separation of the center
of the $i$th and $j$th chromonic molecules,
${\bf pu}_i^{\rm cc}$ and ${\bf u}_i^{\rm cc}$ are a unit vectors lying 
parallel and perpendicular to the plane of the 
the $i$th chromonic molecule, 
$P_2(x) = 3/2\cos^2(x) -1/2$ is the second Legendre polynomial, and the sums range
over all pairs of molecules.
 
In practice, we work either with fully angle-averaged correlation functions, e.g.
\begin{equation}
g^{\rm cc}(r) = {1 \over {4 \pi}} \int d\Omega\ g^{\rm cc}(r, \Omega),
\end{equation}
or with cylindrically-averaged correlation functions, e.g.
\begin{equation}
g^{\rm cc}(r_\parallel, r_\perp) = {1 \over {2 \pi}}
\int_0^{2 \pi} d\phi\ g^{\rm cc}(r_\parallel, r_\perp, \phi).
\end{equation}
Here, $r_\parallel$ and $r_\perp$ denote the components of ${\bf r}$ parallel
and perpendicular to the orientation of the disc at the origin of the coordinate frame
(see figure \ref{free_geometry}(a)), i.e.
$r_\parallel = {\bf r} \cdot {\bf u}^{\rm cc}$ and
$r_\perp = |{\bf r} - ({\bf r} \cdot {\bf u}^{\rm cc}) {\bf u}^{\rm cc}|$,
and $\phi$ is the azimuthal orientation of ${\bf r}$ about
${\bf u}^{\rm cc}$. $g_1^{\rm cc}(r)$ , $g_2^{\rm cc}(r)$ 
and $g_1^{\rm cc}(r_\parallel, r_\perp)$ are defined in an analogous way.

\begin{figure}[h]
\epsfxsize=6.0 in
\centerline{\epsffile{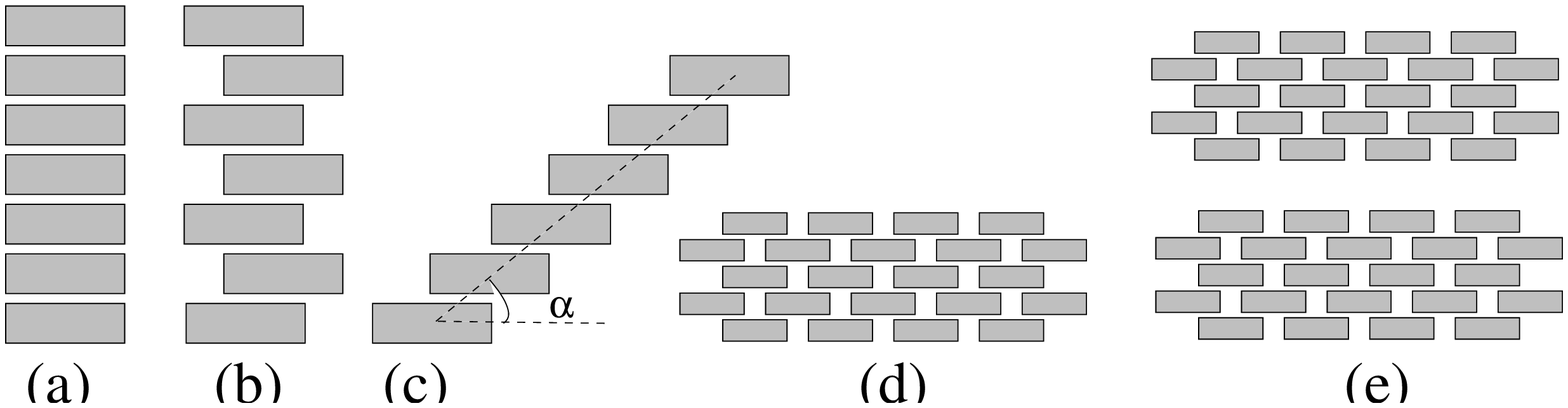}}
\caption{\protect{Schematic representations of chromonic columnar aggregates,
H-aggregates ( a and b), J-aggregates (c), brickwork structure (d) and
layered brickwork (smectic) structure (e) }}
\label{fig1}
\end{figure}

In figure \ref{gr_cor} we have plotted spherically averaged correlation function 
$g^{cc}(r)$ for different chromonic concentration 
and the first peak at $r = 1.12\sigma$ signify strong positional correlation 
between chromonic molecules. 
Successive small peaks at $r = 2.2\sigma, 3.3\sigma$ etc. indicate that 
the molecules aggregate in a columnar fashion as in figure \ref{fig1}(a) or (b).

The pair correlation between water molecules plotted in figure \ref{gss_cor} exhibits
the expected liquid like behavior with short range positional order.

More detailed structural informations for the chromonic aggregate emerges 
from the two-dimensional correlation function $g^{\rm cc}(r_\parallel, r_\perp)$
and $g_1^{\rm cc}(r_\parallel, r_\perp)$, shown in figure \ref{orient_cor}. With the 
reference chromonic molecule situated at the origin
the three prominent peaks at $r_\perp \sim  1.1, 2.2,3.3$ indicate that the
chromonic molecules form columnar aggregates and the molecules within the columns 
is organized as in figure \ref{fig1}(a). Due to close packing effect 
the center of mass of the chromonic molecules within the columns 
can be shifted with respect to 
another by $\sigma$ which is reflected in the broadening of the peaks around 
$r_\parallel \pm  1.0$. Examination
of $g_1^{\rm cc}(r_\parallel, r_\perp)$ reveals a strong intensity ($\sim 1.0$)
peak around $r_\parallel \pm 1.0$ which indicates  that the chromonic molecules 
within the columns are parallel to each other.

We also compute average the aggregation number to study its dependence on 
chromonic concentration.
We identify two chromonic molecules belonging to a cluster 
if the distance separating two monomers belonging to two chromonic molecules
is $\le 2.5 \sigma$. From the cluster size distribution we have calculated 
the average aggregation number through the following equation:

\begin{equation}
<L> = \sum s^2 N(s)/\sum s N(s)
\end{equation}
where $N(s)$ is the number of cluster of size $s$.
In figure \ref{clust} we have shown the average aggregate size $<L>$ as a function of 
chromonic concentration. At low concentration $<L> \sim \phi^{0.6}$. At higher concentration
the dependence of $<L>$ on $\phi$ is stronger than power law. Similar behavior is
observed for conventional lyotropic systems \cite{oligo,yanik99}. 

Since in our simulation we do not observe the formation of regular columnar 
aggregates as shown in figure \ref{fig1}, we have simulated pre-assembled
single and multiple columnar aggregates to test the stability of columnar 
aggregates. The simulation was done in $NVT$ ensemble. The columnar aggregates
remain stable over simulation time, in contrast to model 2 in which case
pre-assembled columnar aggregates break and the system becomes
isotropic. 
In addition our simulation with model 2 do not exhibits the spontaneous formation of 
columnar aggregates (figure \ref{modelasnap}). 
This indicates that we need to make judicious choice 
of the ratio of number of hydrophilic and hydrophobic units, an 
overall hydrophobicity of the chromonic molecules being necessary for the
spontaneous self-assembly.

\section{Free Energy computation by Umbrella Sampling}
To calculate the free energy associated with the formation of columnar aggregates
we have calculated the free energy of association in pulling apart two chromonic
molecules by using umbrella sampling ~\cite{torrie}. 
The distance $r_c$ separating two chromonic molecules 
were confined (figure \ref{free_geometry}) by means of a harmonic potential of the 
form :
\begin{equation}
U(r) = \frac{1}{2} k (r -r_c)^2
\end{equation}
where $r_c$ is the equilibrium distance imposed between the center of mass of two 
chromonic molecules. The equilibrium distance were varied from $1.1 \sigma$
to $7 \sigma$ in steps of $0.1\sigma$ - $0.2\sigma$. Note that the equilibrium 
distance between two chromonic molecules can not be less than $\sigma$. 
From the individual biased average probabilities, $P(r)$, 
obtained for each $r_c$, unbiased free energy ($F$) can be constructed 
self-consistently using the weighted histogram analysis method (WHAM)
~\cite{frenkel,swendsen}. We have 
performed MC simulation at constant pressure for 25 windows. The value of
the spring constant used was $k = 200$. We have performed some simulation with higher
values of spring constant $k = 2000$ and found qualitatively similar free energy
behavior.  
The probability distribution of the distance $r$ separating two chromonic 
molecules is ,
\begin{eqnarray}
P(r) & = & \langle \delta(r - r_c) \rangle \nonumber \\
   & = & {1 \over Z} \int d{\bf r}^N \delta[r - r_c)]
   \exp [-\beta V({\bf r}^N)],
\end{eqnarray}
for values of $r$ for which $P(r)$ is exceedingly small. Here, $\delta$
is a Dirac delta function, $N$ is the number of particles,
${\bf r}^N$ denotes the set of particle coordinates, $\beta = 1 / (k_B T)$,
where
$k_B$ is Boltzmann's constant and $T$ is the absolute temperature,
$V({\bf r}^N)$ is the potential energy, and $Z$ is a normalization
factor (the configurational partition function).
However, problem arises due to the fact that 
to $P(r)$ becomes exceedingly small for values of $r$ which give significant 
contribution to the free energy. 
Umbrella sampling makes use of a biasing potential to sample the region of phase space 
for which $P(r)$ is exceedingly small.
The distribution
of $r$ in the presence of a biasing potential is
\begin{eqnarray}
P^\prime(r) & = & {1 \over {Z^\prime}} \int d{\bf r}^N \delta[r - r_c)] \nonumber\\
  & & \exp \{-\beta [V({\bf r}^N) + U(r)]\} \nonumber \\
   & = & {Z \over {Z^\prime}} \exp[-\beta U(r)] P(r),
\end{eqnarray}
where $Z^\prime$ is the partition function for the biased Hamiltonian.
From this it follows that
\begin{equation}
P(r) = {{Z^\prime} \over Z} \exp[\beta U(r)] P^\prime(r).
\end{equation}

Thus, the distribution function $P(r)$ can be obtained (to within
a multiplicative constant) from a measurement of the biased
distribution $P^\prime(r)$. The Helmholtz free energy $F$ as a function of
$r$ can then be obtained (to within an additive constant) from
\begin{equation}
F(r) = - k_B T \ln[P(r)].
\end{equation}
By piecing together the relative free energies measured using a number of
biasing potentials, it is possible to construct $F(r)$ over any specified
range of $r$. 

We have computed the potential of mean force of bringing two 
initially separated chromonic molecules
on top of each other by the above method and the result is 
shown in figure \ref{free_energy2}. 
This case is referred to column 2 (figure 
\ref{free_geometry}(a)). From the free energy curve we see that there is 
strong attraction between chromonic molecules at short distance, leading to face-to-face
chromonic aggregation. Both energetic and entropic factors lead to this attraction: 
by stacking on top of each other the unfavorable hydrophobic interaction with water 
molecules is reduced and this stacking results in a net increase in the 
volume available to the water molecules and thereby increasing their entropy.
At intermediate separation water molecules form solvation shells between
the two chromonic molecules which results in a depletion repulsion between 
the chromonic molecules and thereby increasing the free energy barrier between them.
At larger separation two chromonics do not interact with each other and so the
potential of mean force levels off. However, at this stage we are unable to 
separate the entropic and energetic contribution in the free energy.

We have also computed different orientational 
pair correlations $g_1^{cc}(r)$ and $g_2^{cc}(r)$ 
between the two chromonic molecules and are shown in figure \ref{orient_cor2}.

From the $g_1^{cc}(r)$ we see that at the minimum separation 
($r = 1.0826\sigma$) the two chromonic molecules are parallel and stacked 
together. However, their in-plane unit vectors make an angle to each other
as is evident from $g_2^{cc}(r)$. As the separation increases they start loosing their
orientational correlations. 
 
Next we compute free energy of association of a chromonic molecules onto a 
columnar aggregates formed by two chromonics (referred to as 
column 3, figure \ref{free_geometry}(b))
 and three chromonic molecules (referred to as column 4, figure \ref{free_geometry}(c)) 
respectively. Within the columnar aggregate, 
the separation and orientation of the two chromonics are kept fixed which was 
computed at the location of the free energy minima for the column 2 case (see figure
\ref{free_energy2}). The computed free energy is shown in figure \ref{free_energy3}.

The free energy curves has the same features as in the case column 2 (figure 
\ref{free_energy2}). At large separation chromonic molecules do not see each other and
consequently free energy reaches a plateau at larger separation. At intermediate 
separation the chromonic molecules start interacting with each other and start
expelling the water molecules form the region between them to aggregate on top
of each other and this leaves more room for the water molecules. This way water 
molecules gain entropy and chromonic molecules gain energetically (less contact
with water molecules). This depletion effect eventually leads single chromonic
molecules going on top of the existing columnar aggregates of length 2 or 3. Such type
of depletion attraction and repulsion has been found between macromolecules
immersed in a fluid of much smaller particles \cite{asakura,yodh,biben}.
As in the case of column 2, we have also computed the orientational pair correlation
for column 3 and 4. In case of column 3, the behavior is similar 
to the case of column 2. However for column 4, 
now the chromonic molecules remain perfectly parallel both within and perpendicular
to the chromonic plane. The presence of three chromonic molecules within the aggregate
strongly influence the fourth one whcih is coming on top of it.

\section{Summary and Conclusion}
We have developed a class of idealized model for the chromonic molecules. 
Simple site-interaction models represent a useful class of models for the 
study of the chromonic liquid crystalline self-assembly. For sufficiently 
hydrophobic molecular models formation of columnar aggregates is observed. 
Our free energy computation for different length of columnar aggregates 
gives valuable insight. However, more effort is required to test the isodesmic
hypothesis. Future work will focus on improving configurational sampling (via
rigid-body molecular dynamics) and on investigations of chemically realistic 
models of chromonic LCs.

\section{Acknowledgements}
The work is supported by NSF MRSEC Grant DMR 98-09555.

\newpage

\newpage
\begin{figure}[h]
\centerline{\epsffile{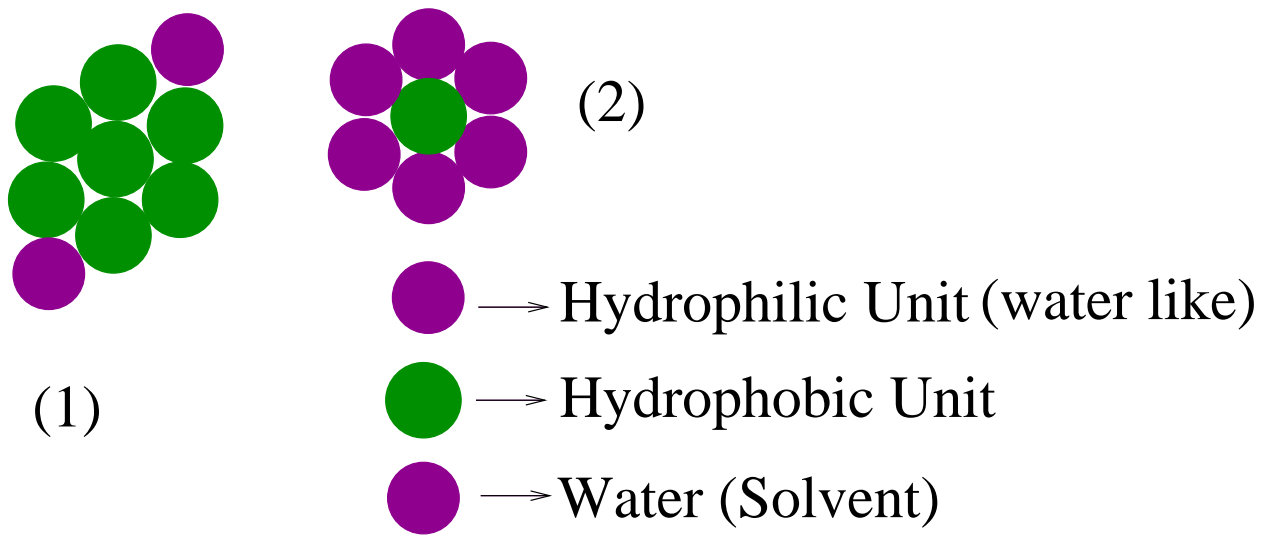}}
\caption{\protect{ A schematic representation of model chromonic 
and water molecules.}}
\label{fig2}
\end{figure}
\newpage

\begin{figure}[h]
\centerline{\epsffile{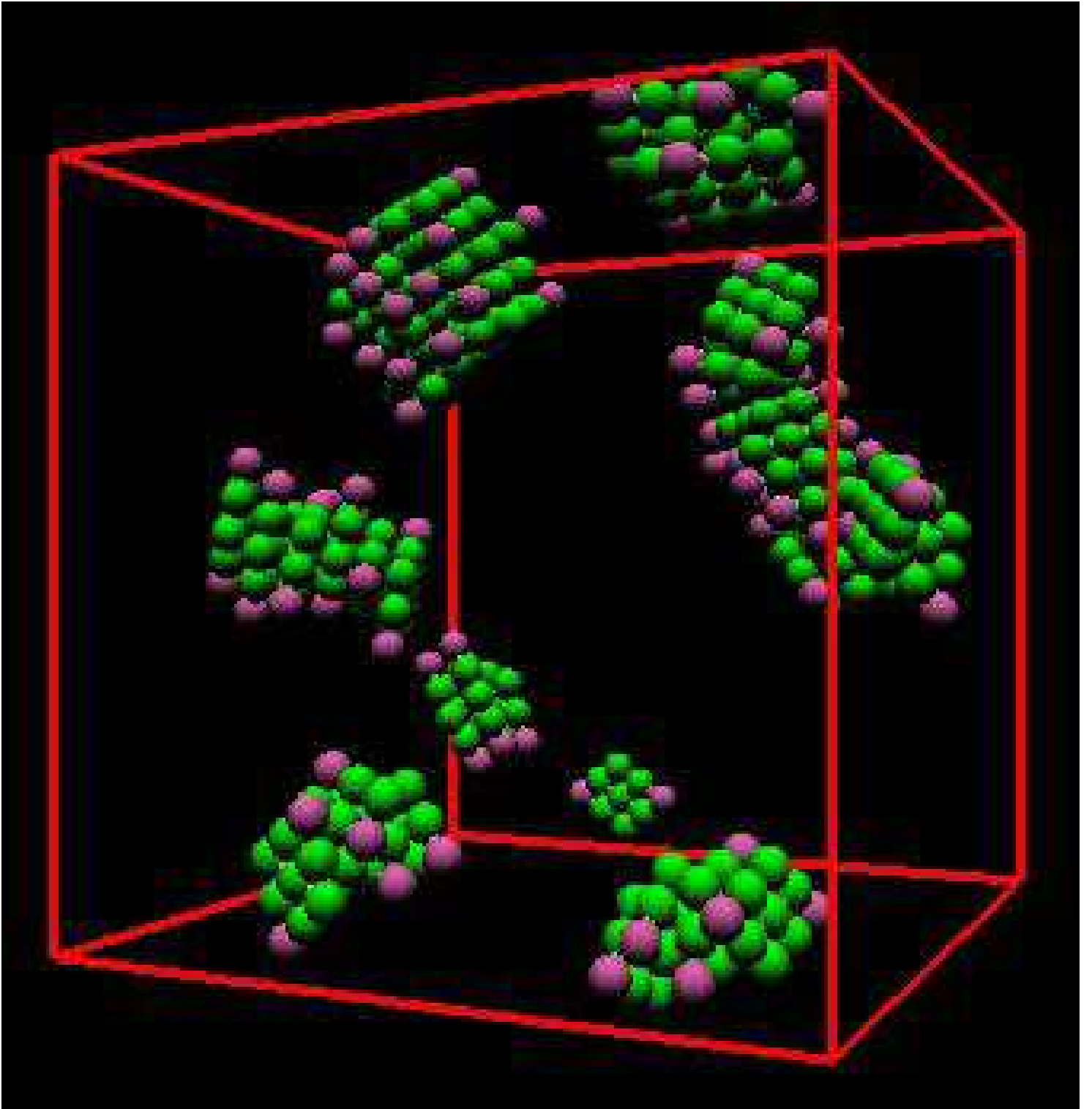}}
\caption{\protect{ Starting from an isotropic initial condition 
we get columnar aggregates. The parameters are: $\phi = 0.081$, $N_c$ = 50, 
$N_W$ = 5100, $T^* = 1.0$ $P^* = 1.0$, all in reduced units. For clarity 
the water molecules have not been shown.}}
\label{snap1}
\end{figure}

\newpage
\begin{figure}[h]
\centerline{\epsffile{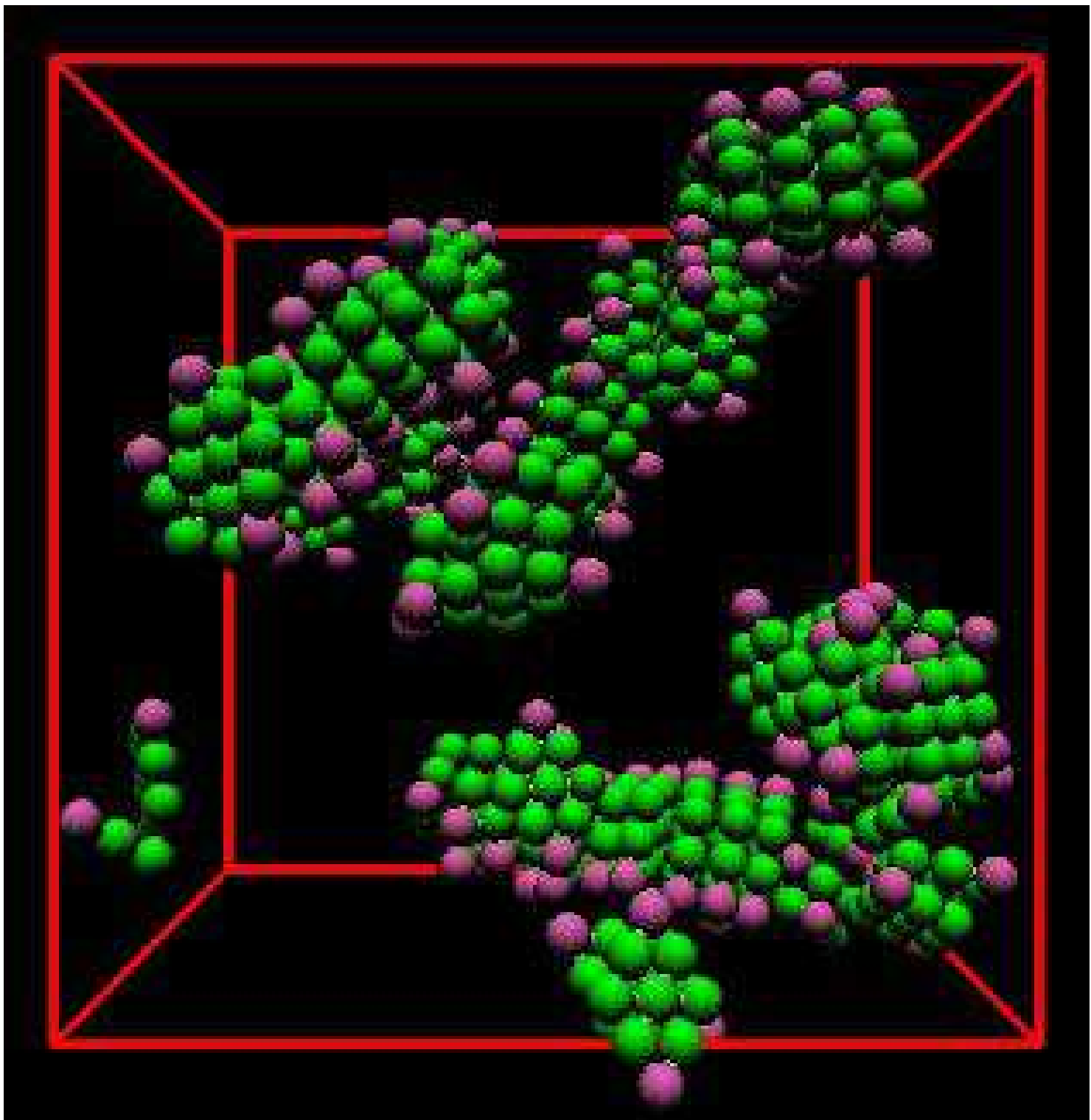}}
\caption{\protect{ Same as in figure \ref{snap1}. The parameters are: 
$\phi = 0.137$, $N_c$ = 90, $N_W$ = 5100, $T^* = 1.0$ $P^* = 1.0$, all in reduced units. 
For clarity the water molecules have not been shown.}}
\label{snap2}
\end{figure}

\newpage
\begin{figure}[h]
\centerline{\epsffile{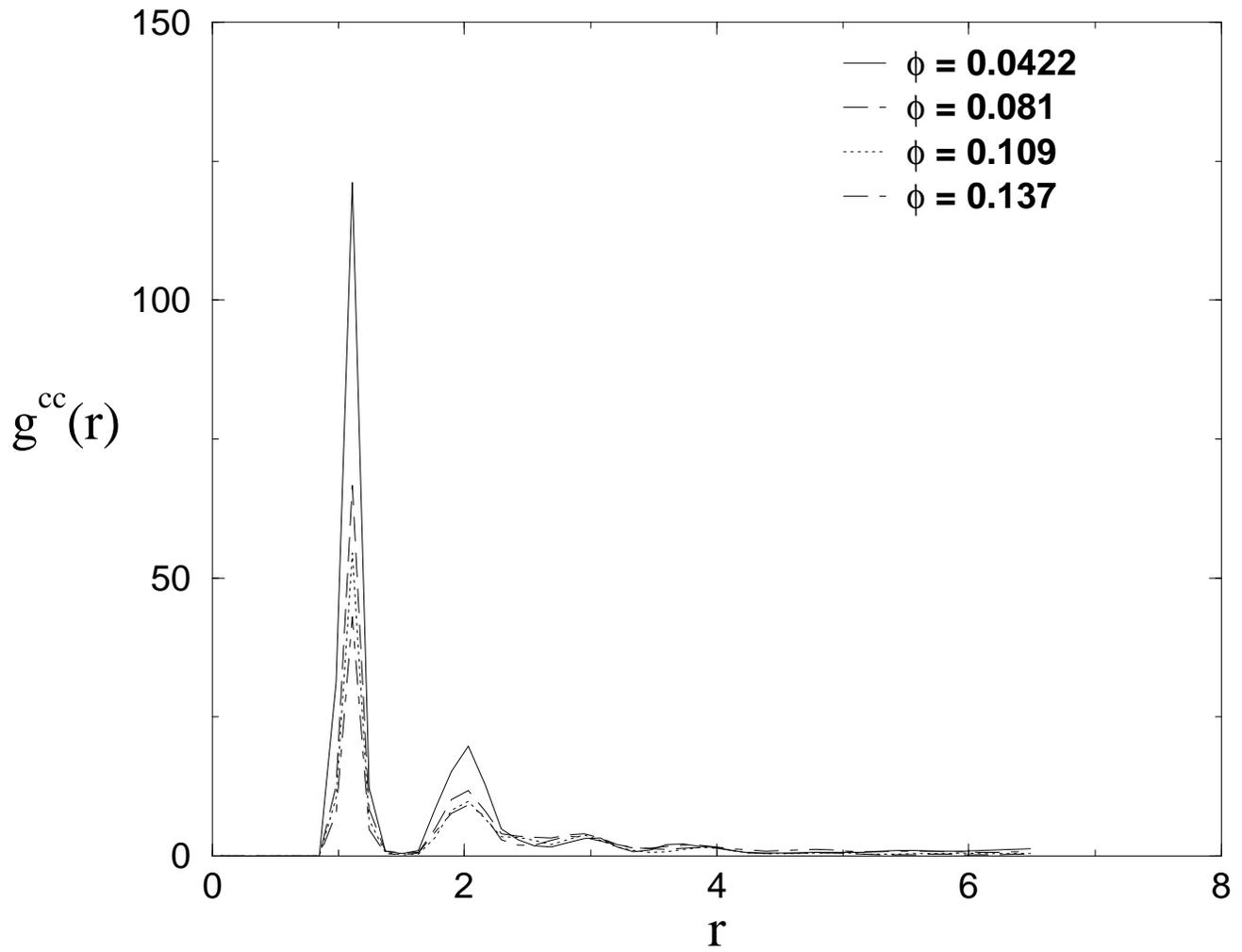}}
\caption{\protect{Pair correlation between the center of mass of the chromonic molecules
for different chromonic concentration.}}
\label{gr_cor}
\end{figure}

\newpage
\begin{figure}[h]
\centerline{\epsffile{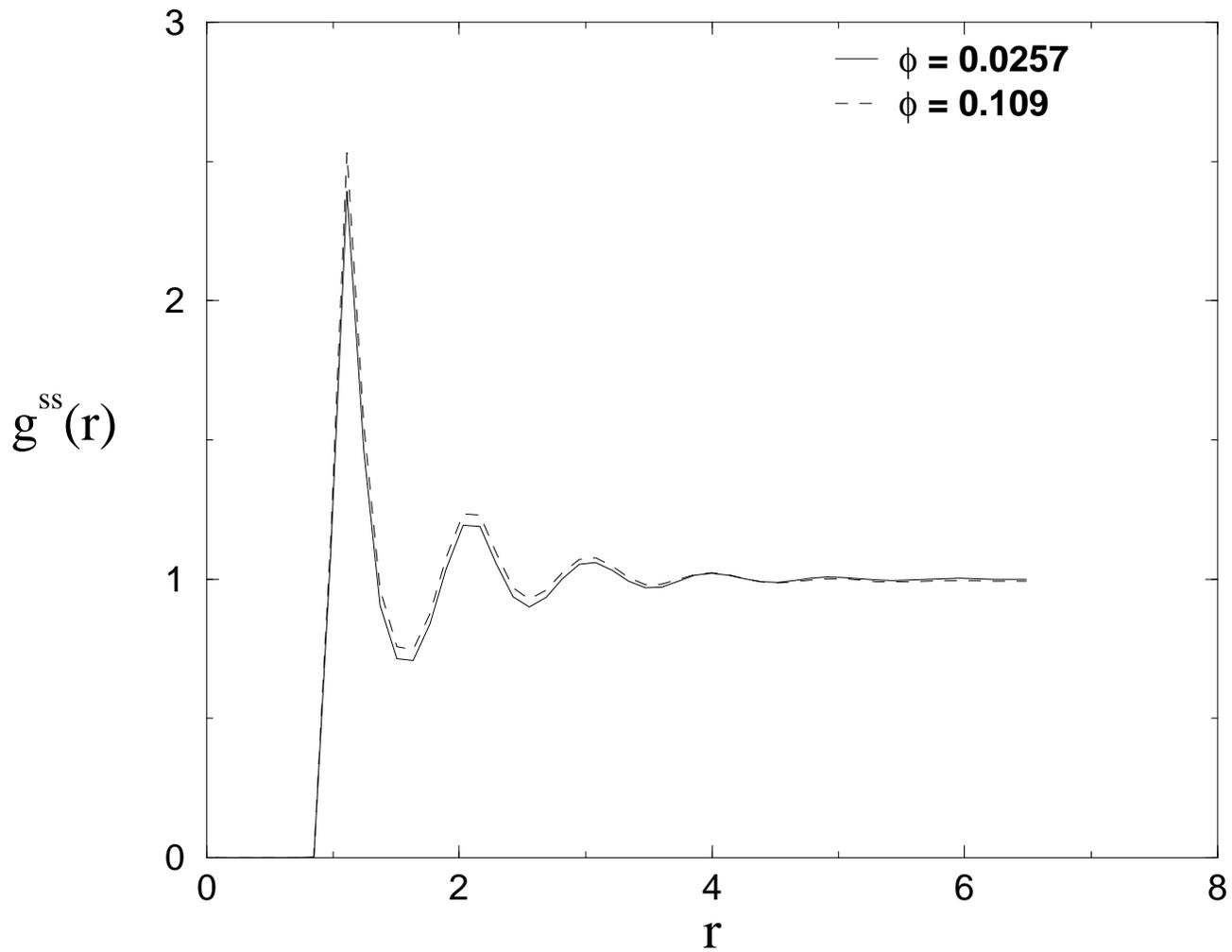}}
\caption{\protect{Pair correlation for water molecules
for different chromonic concentration.}}
\label{gss_cor}
\end{figure}

\newpage
\begin{figure}[h]
\centerline{\epsffile{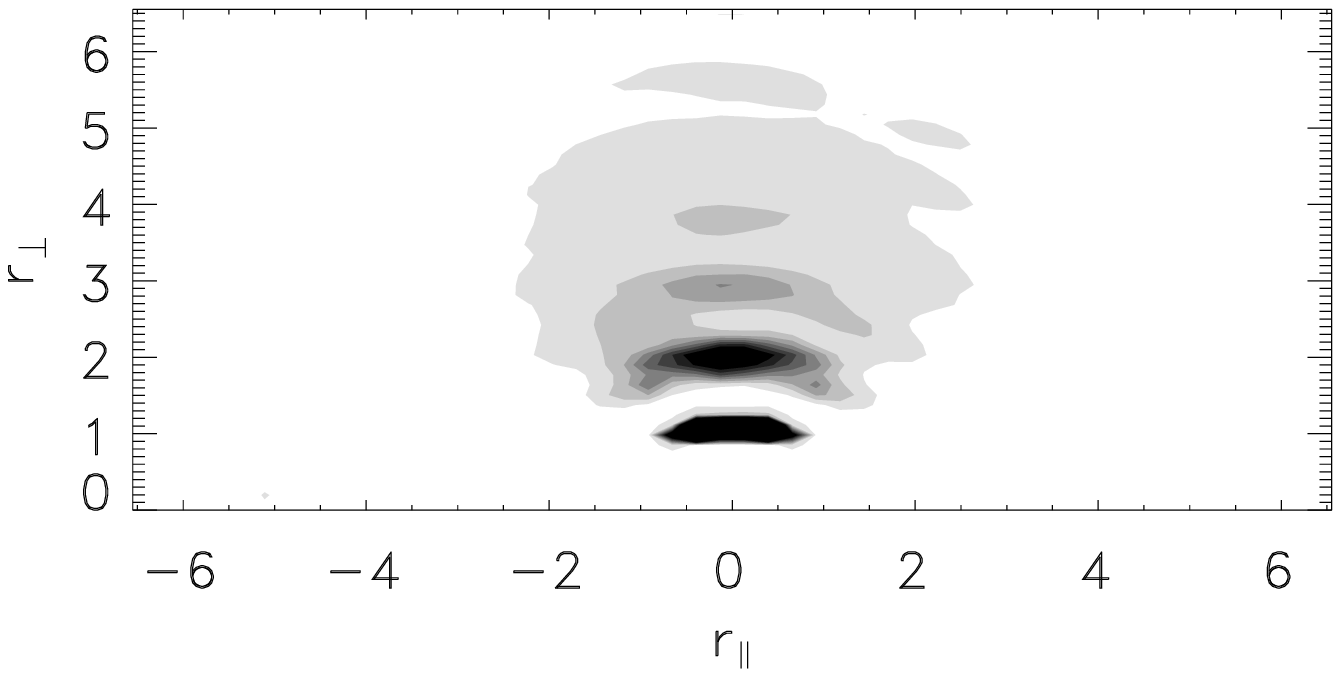}}
\centerline{\epsffile{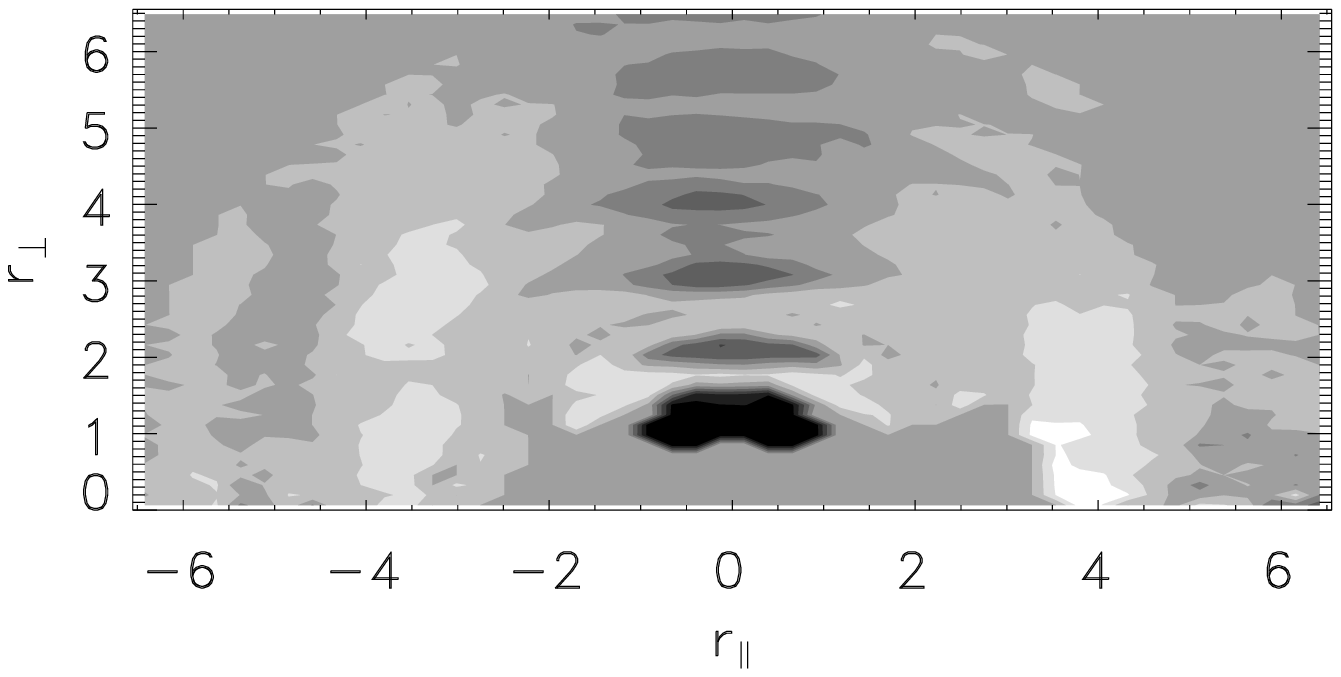}}
\caption{\protect{Two dimensional correlation between the center of mass of the chromonic 
molecules (a) $g^{\rm cc}(r_\parallel, r_\perp)$ and 
(b) $g_1^{\rm cc}(r_\parallel, r_\perp)$. The parameters are same as in figure 
\ref{snap2}. }}
\label{orient_cor}
\end{figure}

\newpage
\begin{figure}[h]
\centerline{\epsffile{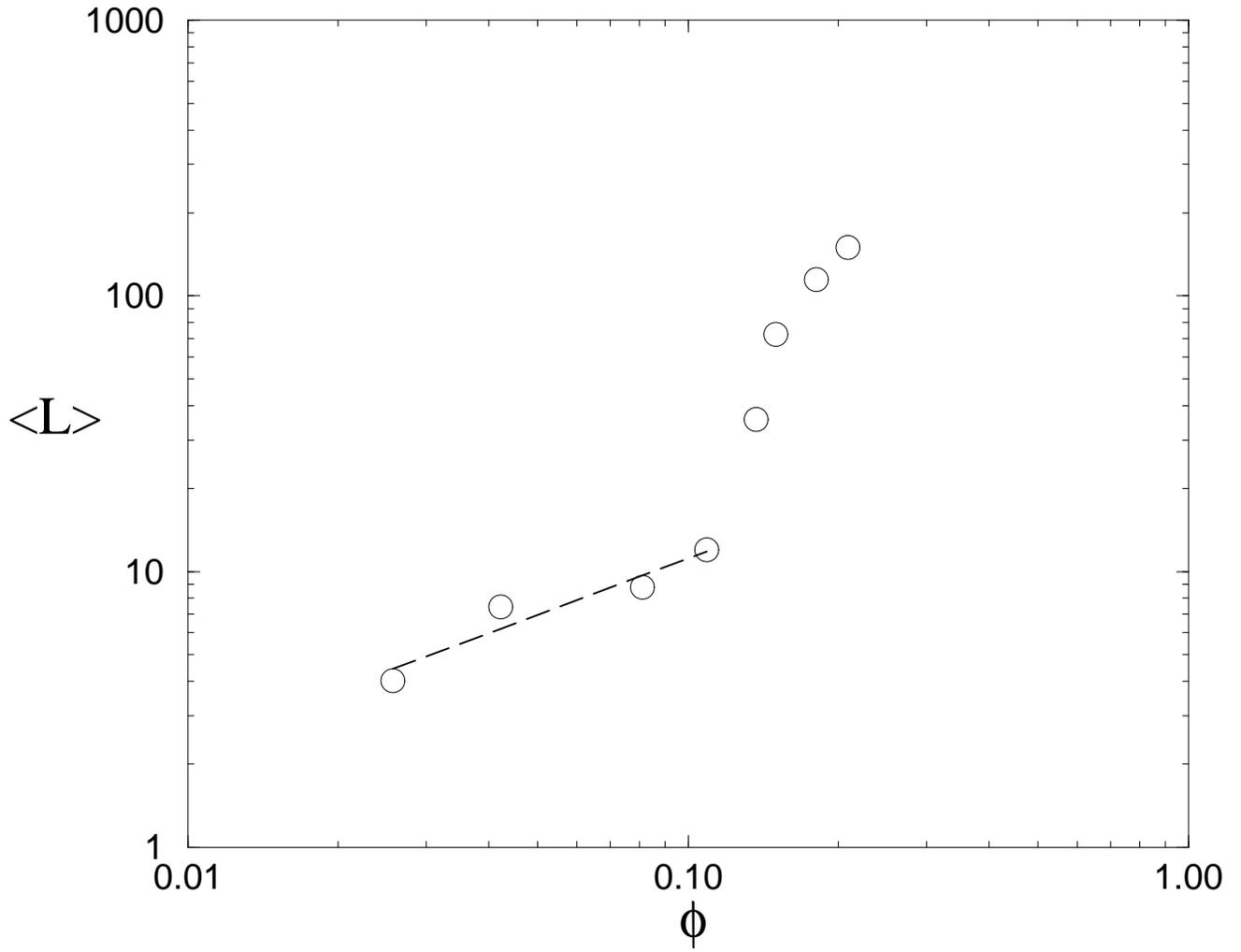}}
\caption{\protect{Average aggregation number $<L>$ as a function of chromonic 
concentration $\phi$.}}
\label{clust}
\end{figure}

\newpage
\begin{figure}[h]
\centerline{\epsffile{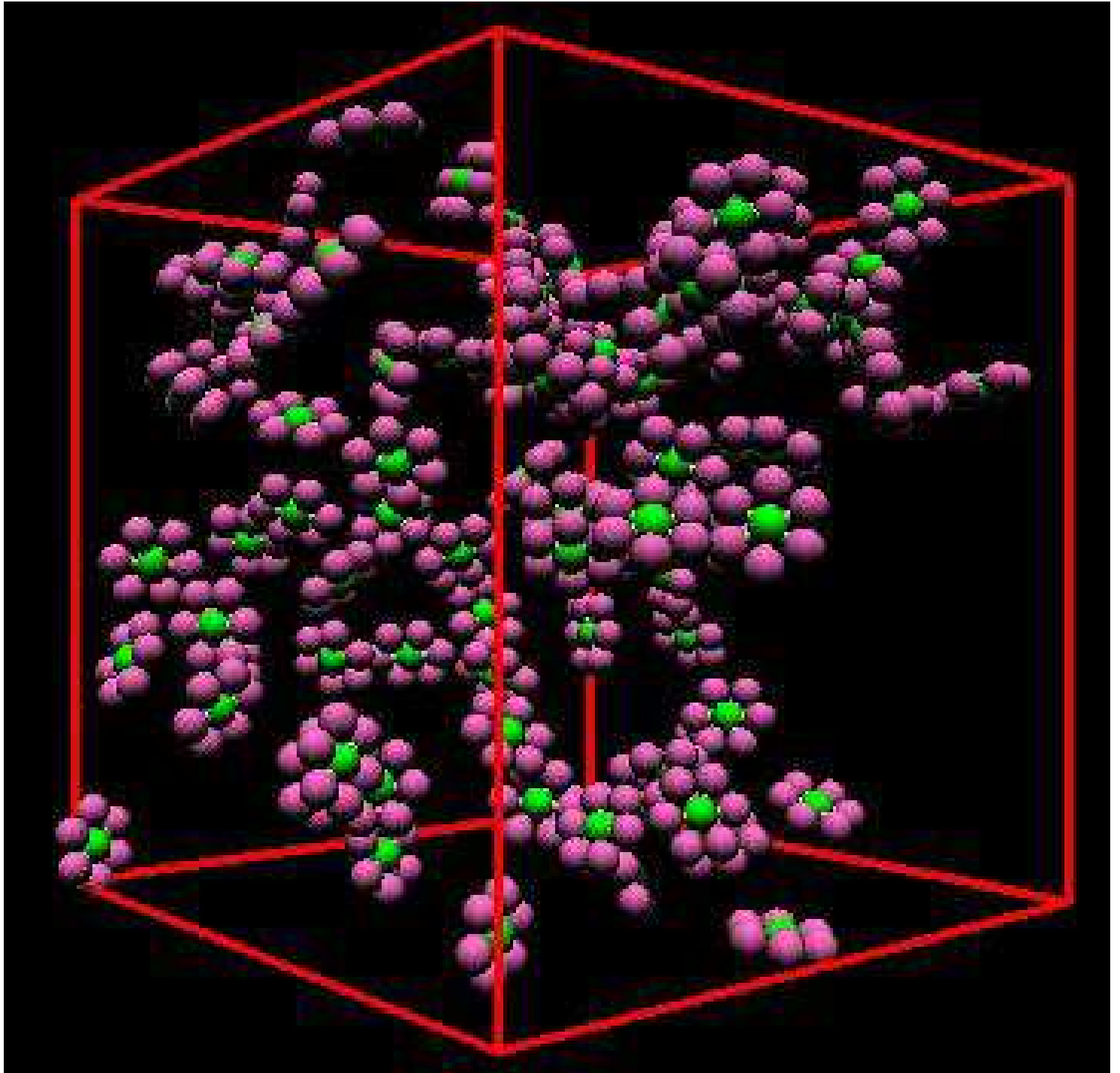}}
\caption{\protect{Simulation with model 2 does not give rise to columnar aggregates.
The chromonic density is $\phi = 0.12$. Other parameters are the same as in figure 
\ref{snap1}.}}
\label{modelasnap}
\end{figure}

\newpage
\begin{figure}[h]
\centerline{\epsffile{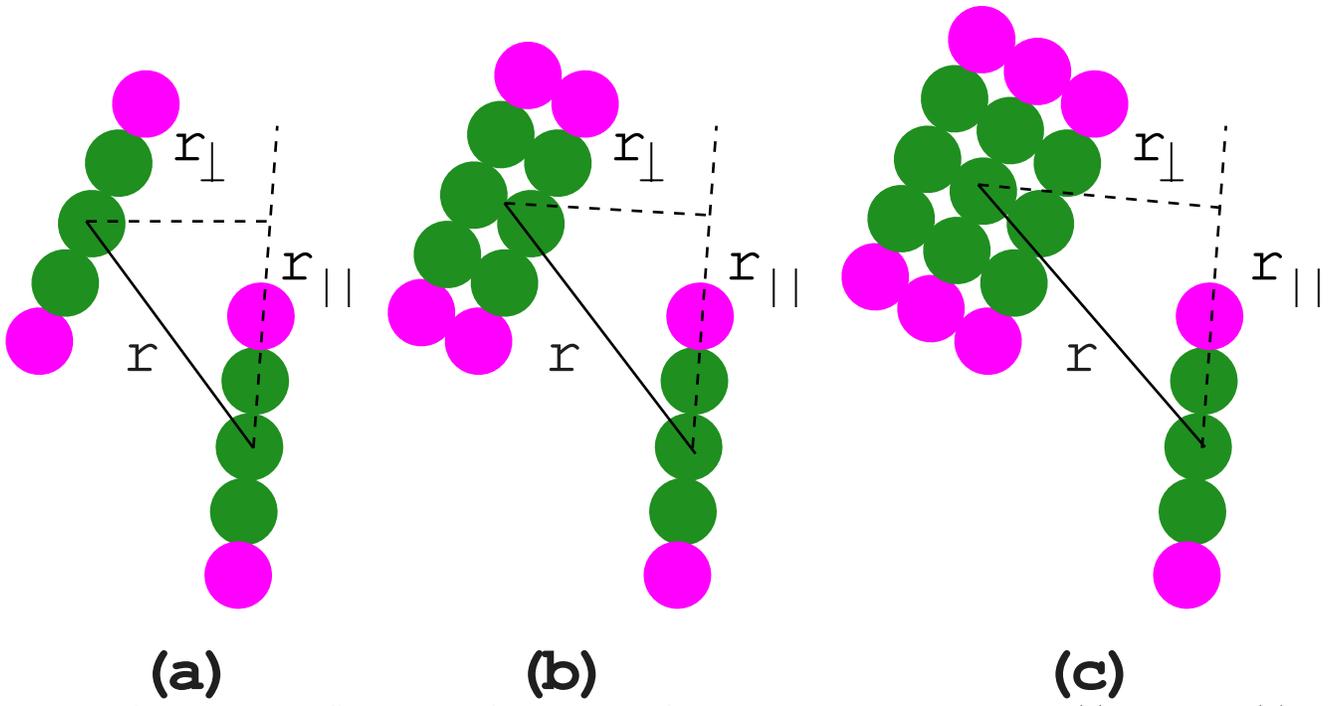}}
\caption{\protect{Three different cases for which the free energy has been computed.
(a) column 2, (b) column 3 and (c) column 4.}}
\label{free_geometry}
\end{figure}

\newpage
\begin{figure}[h]
\centerline{\epsffile{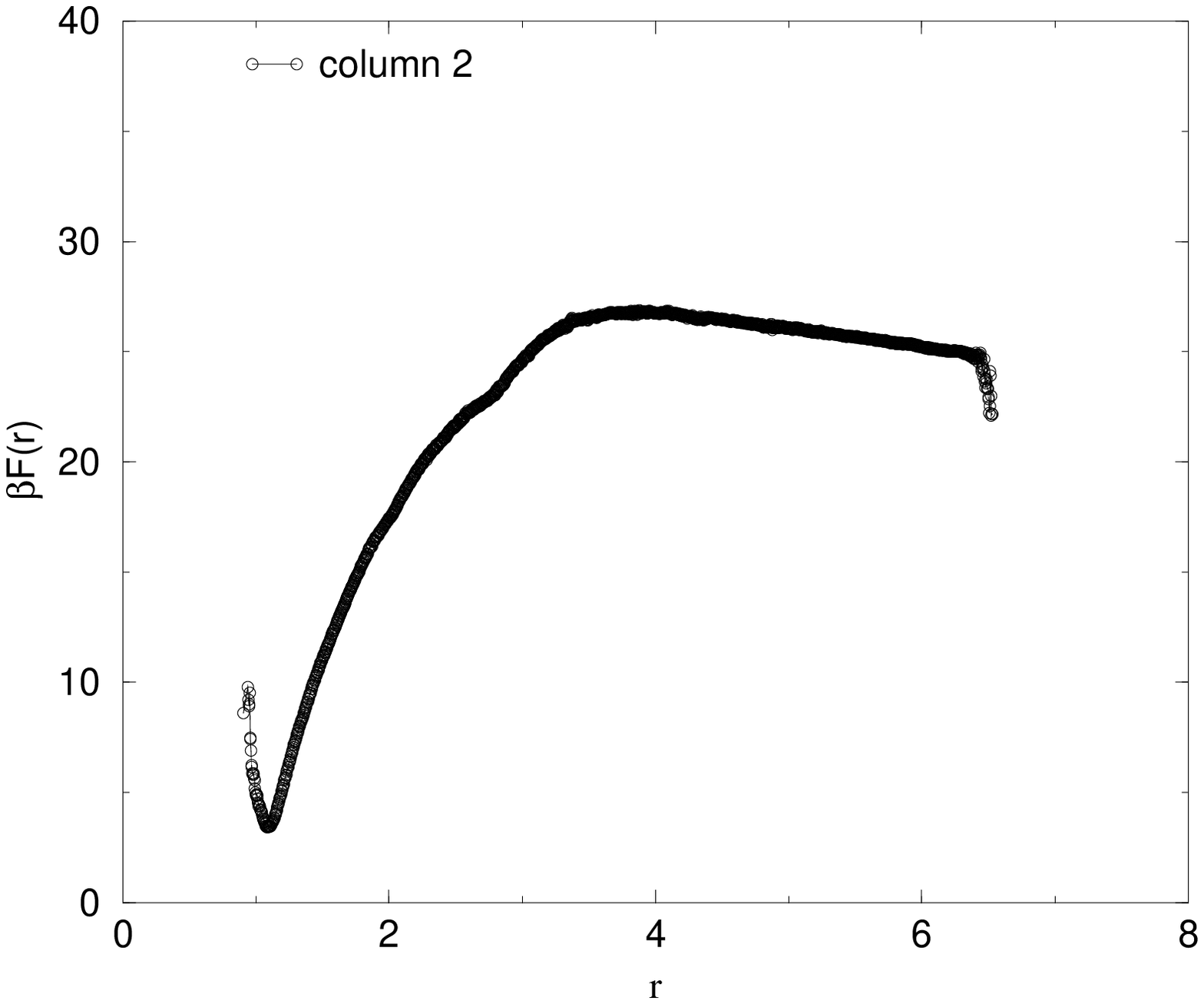}}
\caption{\protect{Free energy as a function of the distance separating the center-of-mass
of two chromonic molecules for the case \ref{free_geometry}(a).}}
\label{free_energy2}
\end{figure}

\newpage
\begin{figure}[h]
\centerline{\epsffile{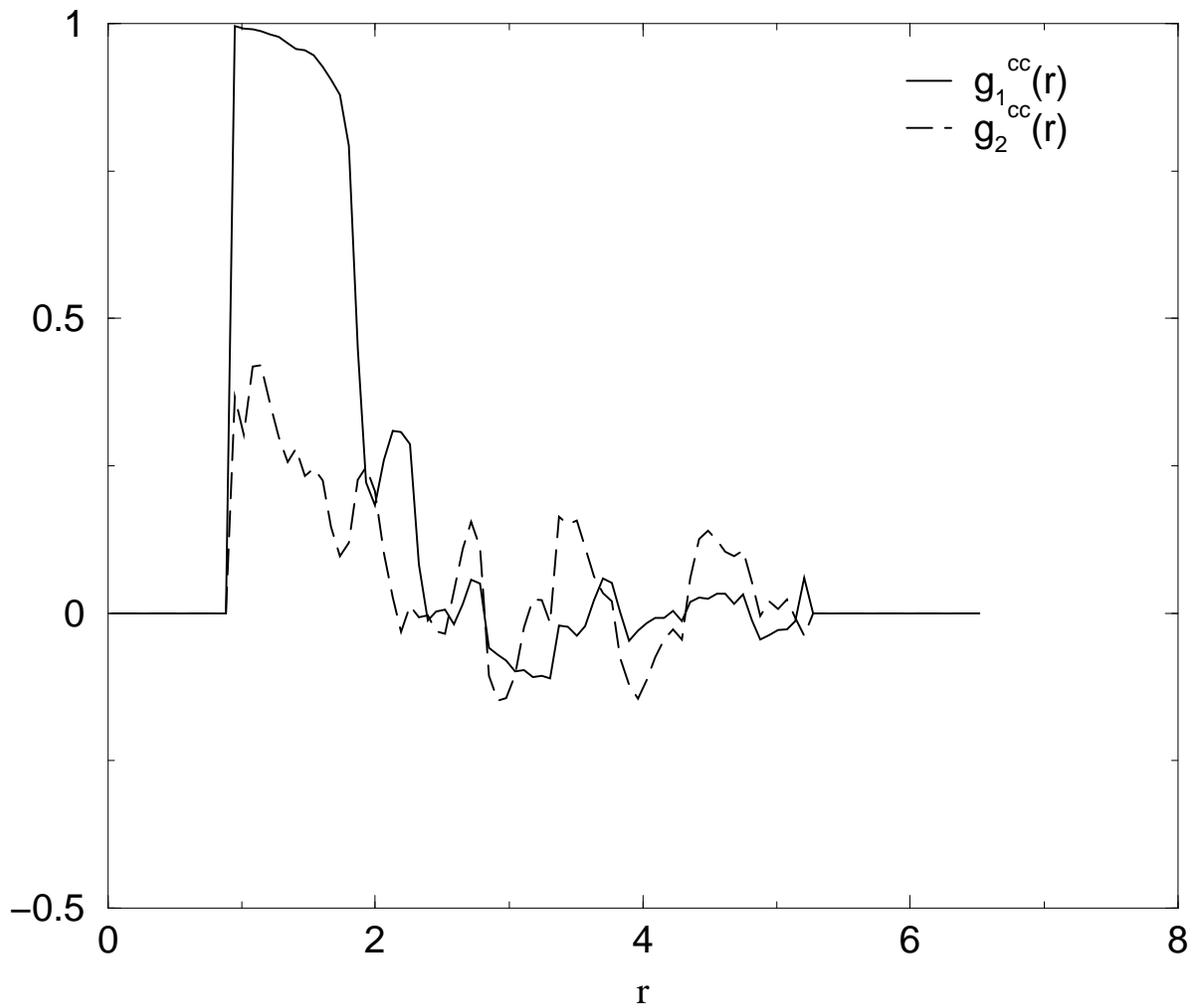}}
\caption{\protect{Orientational pair correlations between two chromonic molecules for
the case \ref{free_geometry}(a).}}
\label{orient_cor2}
\end{figure}

\newpage
\begin{figure}[h]
\centerline{\epsffile{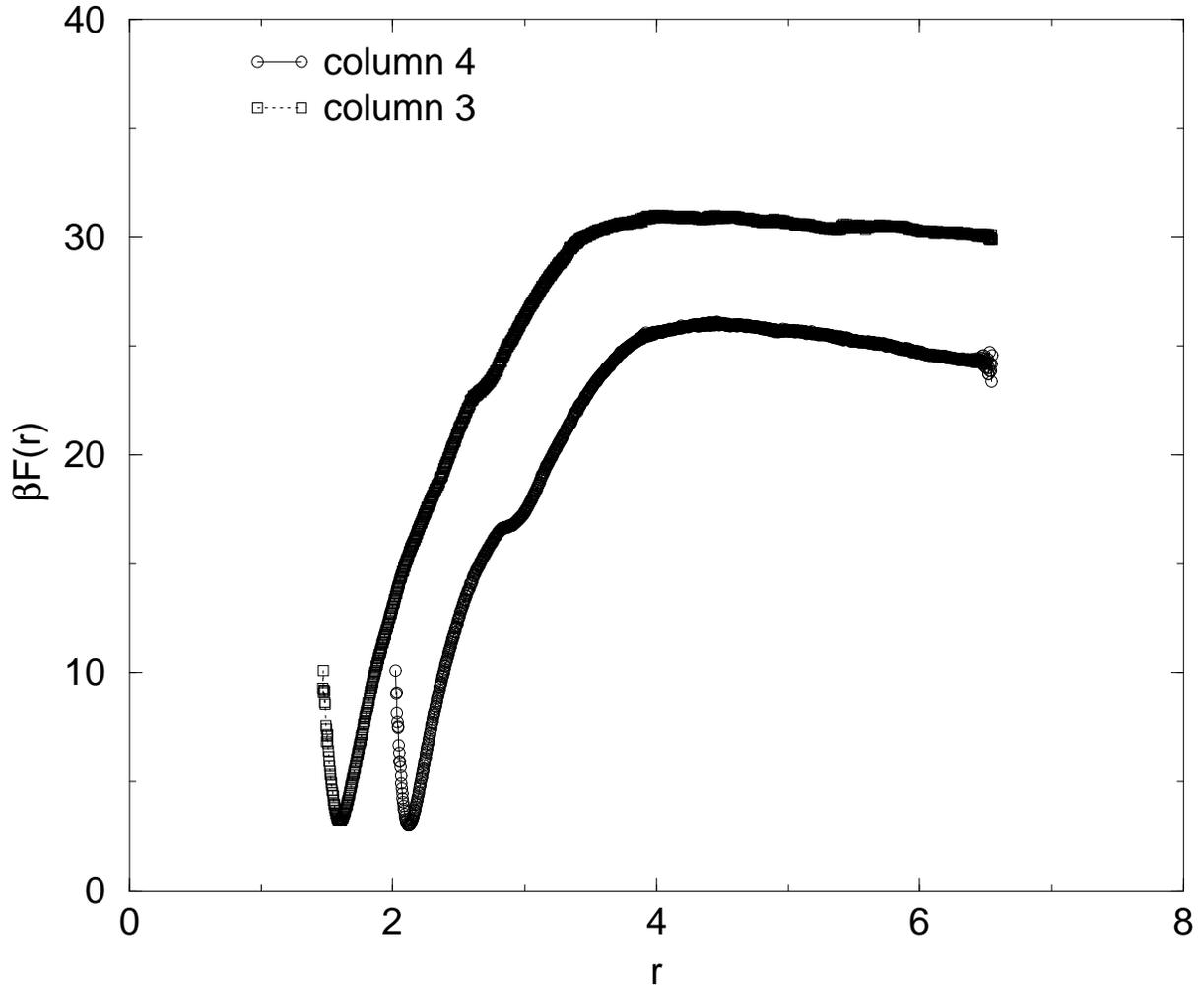}}
\caption{\protect{Free energy as a function of the distance separating the center-of-mass
of two chromonic molecules for the case shown in figure \ref{free_geometry}(b) and (c).}}
\label{free_energy3}
\end{figure}

\newpage
\begin{figure}[h]
\centerline{\epsffile{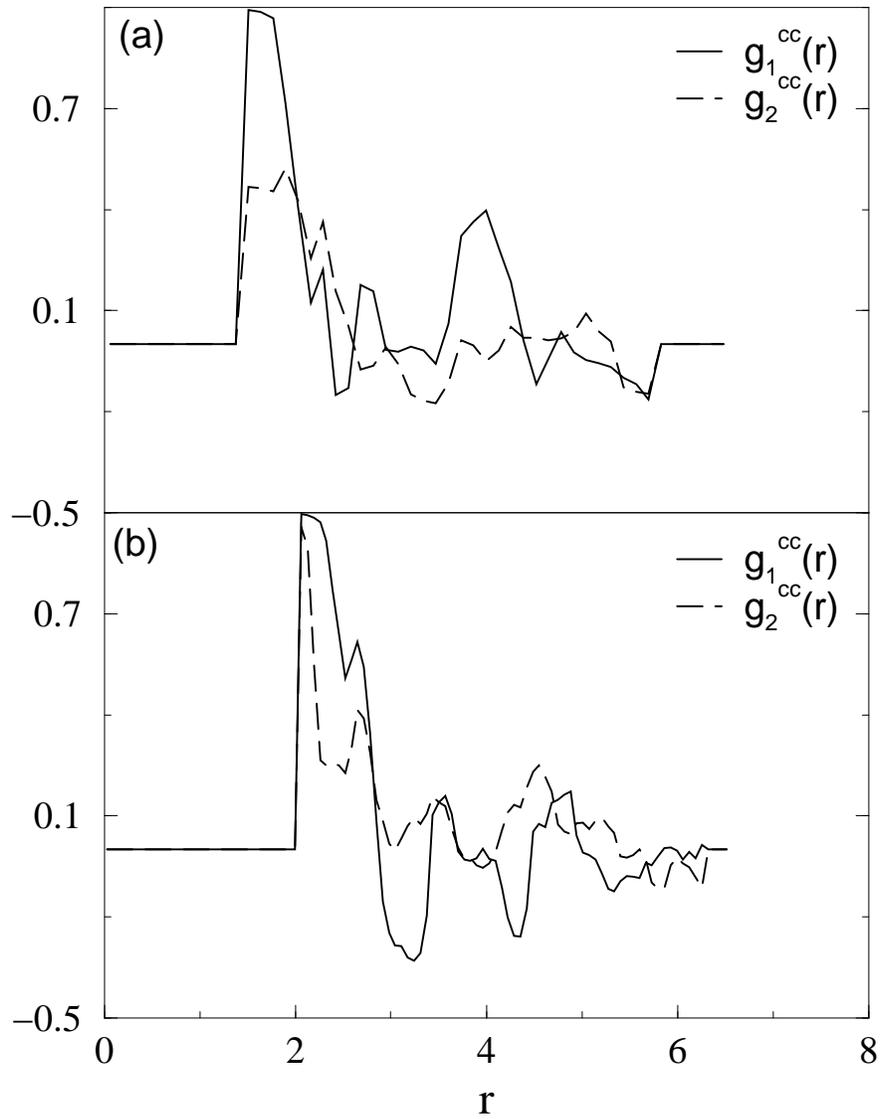}}
\caption{\protect{Orientational pair correlations between chromonic molecules (a) for
the case shown in figure \ref{free_geometry}(b) and (b) for the case shown in figure
\ref{free_geometry}(c).}}
\label{orient_cor3}
\end{figure}
\end{document}